\documentclass{eas}
\usepackage{graphicx}

\newcommand{\bea}{\begin{eqnarray}}
\newcommand{\eea}{\end{eqnarray}}
\newcommand{\be}{\begin{equation}}
\newcommand{\ee}{\end{equation}}

\begin{document}

\title{A numerical study of Penrose-like inequalities
in a family of axially symmetric initial data } 
\newcommand*{\AEI}{Max-Planck-Institut f\"ur Gravitationsphysik,
  Albert-Einstein-Institut, Am M\"uhlenberg 1, D-14476 Golm, Germany}
\newcommand*{\MEU}{Laboratoire de l'Univers et de ses Th\'eories, UMR 8102
  du C.N.R.S., Observatoire de Paris, F-92195 Meudon Cedex, France}
\newcommand*{\IAA}{Instituto de Astrof\'{\i}sica de Andaluc\'{\i}a, CSIC, Apartado Postal 3004, Granada 
        18080, Spain} 

\author{Jos\'e Luis Jaramillo}\address{\IAA}\secondaddress{\MEU}
\author{Nicolas Vasset}\sameaddress{2}
\author{Marcus Ansorg}\address{\AEI}
\begin{abstract}
Our current picture of black hole gravitational collapse
relies on two  assumptions: i) the resulting
singularity is hidden behind an event horizon
--- weak cosmic censorship conjecture --- and ii)  spacetime
eventually settles down to a stationarity state.
In this setting,
it follows that the minimal area containing an apparent horizon
is bound by the 
square of the total ADM mass (Penrose inequality conjecture).
Following Dain {\em et al.} 2002, 
we construct numerically a family
of axisymmetric initial data 
with one or several marginally
trapped surfaces. Penrose and related geometric inequalities
are discused for these data. As a by-product, it is shown
how Penrose inequality can be used
as a diagnosis for an apparent horizon finder numerical routine.
\end{abstract}
\maketitle
\runningtitle{A numerical study of Penrose-like inequalities}

{\bf  1. Introduction and methodology.}
Our present goal is the study, using numerical techniques, of certain 
geometric inequalities  conjectured to hold
in asymptotically flat Cauchy slices containing an {\em apparent horizon} (AH). 
This is 
an ambitious objective, since numerical tools can offer at best 
a counterexample -- 
not a general proof -- 
and experience with these inequalities has revealed the difficulty of this task. 
Precisely because of this confidence in the inequalities,  the line of reasoning can be reversed
in appropriate settings: inequalities can be (tentatively) taken 
for granted and used as a diagnosis to test specific geometric and/or numerical constructions. 
At the end of the day, we aim at gaining geometric insight about these
inequalities in regimes that are difficult to probe by standard analytic techniques.
We formulate here our strategy and present some preliminary 
results.

{\bf 1.1. Geometric inequalities.}
Penrose inequality is our prototype of geometric inequality.
It follows from a chain of heuristic arguments (Penrose 1973) in
the context of black hole gravitational
collapse, in particular probing {\em weak cosmic censorship conjecture} and the assumption
of an evolution towards a final stationary state.  Penrose inequality
conjectures: $A \leq 16\pi M_{_{\mathrm{ADM}}}^2$, 
where  $ M_{_{\mathrm{ADM}}}$
is the total ADM mass and $A$ is the minimal area enclosing the (possibly non-connected) 
AH. It has been proved in the Riemannian
case $K_{ij}=0$ (Huisken \& Ilmanen 2001, Bray 2001),
an equality is conjectured to hold only for Schwarzschild.
Here we focus on axisymmetric data, for which an angular momentum $J$
can be unambiguously 
defined and Penrose inequality can be strenghtened (Dain et al. 2002) to
\bea
\label{e:Penrose_J}
A \leq 8\pi \left(M_{_{\mathrm{ADM}}}^2 + \sqrt{ M_{_{\mathrm{ADM}}}^4 - J^2}\;\right )
\nonumber \ \ ,
\eea
where equality is conjectured to hold only for Kerr data. We shall refer to this 
latter point 
as the {\em Dain's rigidity conjecture}. Rhs expression 
only makes sense if 
\bea
\label{e:Dain}
|J| \leq M_{_{\mathrm{ADM}}}^2 \nonumber \ \ ,
\eea
an angular momentum-mass inequality recently proved by Dain 
for vacuum, maximal ($K=0$), asymptotically flat, axisymmetric data
with a connected AH (Dain 2007 and references therein).
Equality holds only for extremal Kerr data.
 Petroff and Ansorg have proposed a quasi-local bound for $|J|$ in terms of the area $A$ 
\bea
\label{e:Petroff_Ansorg}
8\pi |J| \leq A  \nonumber \ \ ,
\eea
in the restricted setting of stationary, 
axisymmetric configurations of black holes surrounded by matter. This conjecture 
has been extended to include charges, 
and equality has been shown to exactly correspond to the extremal case (Ansorg
\& Pfister 2007). Simultaneously, it has been argued (Booth \& Fairhurst 2007) 
the non-validity
of this quasi-local inequality for generic  
axisymmetric data. Here we consider 
this latter non-stationary generic situation.
We rewrite previous inequalities in terms of bounded
dimensionless parameters $\epsilon_{_P},\epsilon_{_A},\epsilon_{_D},
\epsilon_{_{PA}}$: 
\be
\begin{array}{lcl}
\epsilon_{_P} := \frac{A}{16\pi M_{_{\mathrm{ADM}}}^2}\leq 1  \ \ &,& \ \
\epsilon_{_D} := \frac{|J|}{M_{_{\mathrm{ADM}}}^2}\leq 1 \\
\epsilon_{_A} := \frac{A}
{8\pi \left(M_{_{\mathrm{ADM}}}^2 + \sqrt{ M_{_{\mathrm{ADM}}}^4 - J^2}\;\right )}\leq 1 \ \ &,& \ \
\epsilon_{_{PA}} := \frac{8\pi |J|}{A} \leq 1 \ \ .
\end{array}\nonumber
\ee
In particular, Dain's rigidity conjecture reads: $\epsilon_{_A}=1 \Leftrightarrow$ 
$(\gamma_{ij},K^{ij})$ are Kerr data.

{\bf 1.2. Axisymmetric Initial Data: deformations on Kerr.} 
A conformal construction of vacuum, maximal and axisymmetric
data --- parametrized 
by two free functions $q$ and $\omega$ ---  is presented in Dain {\em et al.} 2002.
By further restricting $q$ and $\omega$, we have studied: 
i) binary black hole data and ii) deformations of Kerr. We discuss here Kerr deformations and
will present the binary case elsewhere:

1. {\em Choice of $q$}. Fixed by a choice of conformal metric as the 
representative with unit determinant in the conformal class of Kerr in quasi-isotropic
coordinates.

2. {\em Choice of $\omega$} as: $\omega(J,M_{_{\mathrm{Kerr}}},\lambda)= \omega_{_{BY}}(J) - \lambda
 \cdot \omega_{_{\Delta}}(J,M_{_{\mathrm{Kerr}}})$. Here $\omega_{_{BY}}(J)$
is associated with the Bowen-York 
extrinsic curvature in Dain {\em et al.} 2002 method,
whereas $\omega_{_{\Delta}}(J,M_{_{\mathrm{Kerr}}})$  is such that 
$\omega(J,M_{_{\mathrm{Kerr}}},\lambda=1)$ is compatible with Kerr.

3.  {\em Marginally Trapped Outer Surface (MOTS) inner boundary condition}
at an excised sphere of coordinate radius $r=1$, when solving
the Hamiltonian constraint for the conformal factor.
The introduced scale fixes $M_{_{\mathrm{Kerr}}}$ in terms of $J$.

In sum, we work with 
data $[\gamma_{ij}(J, \lambda), K^{ij}(J, \lambda)]$ parametrised
by $J$
and a deformation parameter $\lambda$ ---  Kerr data correspond to $\lambda=1$. 
Data are numerically implemented using both the spectral methods in 
the Meudon C++ Lorene library and the spectral methods developed 
by one of the authors in Ansorg {\em et al.} 2005.

{\bf 1.3. Extraction of geometric information: AH-finders.}
The assessment of the discussed geometric inequalities primarily concerns
AHs properties. First, we need to know their location.
By construction, data in the considered family contain a MOTS at the inner excised
sphere. 
In the generic case, we need an {\em AH-finder} routine to locate
the outermost MOTS. 
In this case, we make use of the spectral 3D {\em spectral integral-iteration}
AH-finder presented in Lin \& Novak 2007.

\begin{figure}
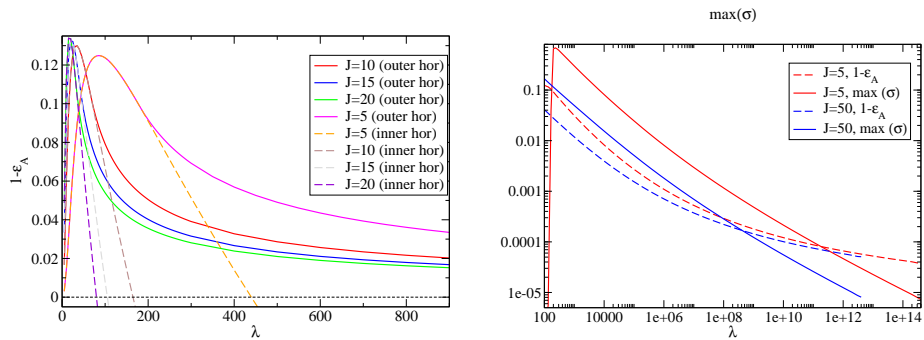

  \centerline{\includegraphics[width=0.47\textwidth]{Figure1.eps}\hspace{1.5em} 
    \includegraphics[width=0.45\textwidth]{Figure2.eps}}
  \caption[]{
    \label{f:Figure1}  
   {\small Left: $(1 - \epsilon_{_A})$ against $\lambda$ showing 
    the emergence of a second horizon, the violation of $\epsilon_{_A}\leq 1$ for the
    inner horizon, whereas Penrose inequality still holds. Right: 
   $(1 - \epsilon_{_A})$ --dashed line-- and the maximum of the outgoing shear $\sigma_+$, 
    --bold line-- against large $\lambda$'s.
}
  }
\end{figure}

{\bf 2. Results.} Fixing  $J$ and screening different values of $\lambda$,
we monitor the dimensionless quantities $\epsilon_{_P},\epsilon_{_A},\epsilon_{_D},
\epsilon_{_{PA}}$. First we check that, as $\lambda$ departs from 
$\lambda=1$, the data indeed move away from Kerr --- i.e. $\epsilon_{_A}$ departs
from $1$, as they should according to Dain's rigidity conjecture.
Increasing $\lambda$,  a critical $\lambda_o$ exists for each $J$ at which 
a second outer horizon {\em detaches} from the inner one.
For sufficiently large $\lambda$, $\epsilon_{_A}$ grows over $1$ for the
inner horizon, whereas the
strenghtend Penrose inequality still holds since the exterior horizon satisfies 
$\epsilon_{_A}\leq 1$ --- cf. Fig. \ref{f:Figure1}. 
The other inequalities are also
satisfied, and no surprises appear. 

At this point we reverse the line of 
reasoning. First, we use the 
inequalities to test the
Lin \& Novak AH-finder. As Fig. \ref{f:Figure1}-left shows, outer
$\epsilon_{_A}$ grows with $\lambda$. Penrose inequality sets a definite
geometric limit to this: $\epsilon_{_A}\leq 1$ should hold 
for all $\lambda$'s if the AH-finder
is properly working. Pushing $\lambda$ to very large values 
(bounded by numerical limits) 
we have checked the validity of the inequality 
--- cf. Fig. \ref{f:Figure1}-right. 
The AH-finder is not producing
spurious solutions and is actually converging to a MOTS, otherwise there is no reason
for $\epsilon_{_A}\leq 1$ to hold. Most importantly, this {\em asymptotic} 
behaviour {\em indicates}
the {\em outermost} character of the outer MOTS.
Additional quantitative tests have
shown (spectral) exponential convergence, an accuracy of $\delta A /A\sim 10^{-12}$, 
and the robustness of the AH-finder. 
Second, Fig. \ref{f:Figure1}-right suggests
$\lim_{\lambda\to\infty} \epsilon_A = 1$. According to Dain's rigidity conjecture
this limit corresponds to Kerr data. 
As a necessary condition for this, the AH should approach a {\em Non-Expanding 
Horizon} (NEH) as $\lambda$ diverges, namely the outer null normal shear $\sigma_+$
should vanish in this limit. 
Fig. \ref{f:Figure1}-right confirms this vanishing asymptotic behaviour, 
reached after an intermediate stage in which the AH definitely departs from a NEH
-- similar results are obtained using a dimensionless $\int_{\cal S} |\sigma_+|^2 dA$.
In this sense, our preliminary numerical results do support Dain's conjecture.

Before concluding, 
we note that a mass-angular momentum
inequality --- milder than Dain's --- 
follows from Penrose and Petroff-Ansorg bounds:
$8\pi|J|\leq A \leq 8\pi  \left(M_{_{\mathrm{ADM}}}^2 + 
\sqrt{ M_{_{\mathrm{ADM}}}^4 - J^2}\;\right )$. 
An {\em optimal fitting} among all three inequalities would be obtained with a 
{\em strong} Petroff-Ansorg inequality:
$8\pi\left( |J| + \sqrt{ M_{_{\mathrm{ADM}}}^4 - J^2} \;\right ) \leq A$. Our numerical
experiments show its violation for $J$ large enough, whereas  
standard Petroff-Ansorg inequality is not disproved. This is indeed a doubtful manner
of proposing new inequalities, that could be referred to as {\em geometric
numerology}.

{\bf 3. Conclusions and perspectives.} We have presented the elements
of an approach to the numerical study of Penrose-like geometric inequalities.
Potential applications to the assessment of numerics  have been illustrated 
by performing a geometric test to the Lin \& Novak AH-finder. More generally, 
Penrose inequality has been proposed as a practical diagnosis in the determination
of the outermost character of a given MOTS. This can be useful in settings where
little intuition is available about the properties of the AH --- 
this has indeed proved to be crucial in our studies of the binary case,
where employed coordinates (Ansorg 2005) strongly affect the form and location
of the AH. Finally, 
following a strategy that resembles a {\em canonical-conjugate}
version of Bray's {\em flow of conformal metrics} approach to 
Penrose inequality (Bray 2001) --- by employing a 
{\em flow of conformal extrinsic curvatures} instead ---
we have found support to Dain's proposal of using $\epsilon_{_A}=1$
as a characterization of Kerr data. Given its cheap cost --- evaluation 
of a single real number --- this has a clear interest for the numerical community.

Next step consists in fully developing the outlined approach. We will pursue
the study of the binary case, much richer and less understood -- e.g.
Dain's mass- angular momentum inequality is not proved in the non-connected case.
We will further assess Dain's rigidity conjecture, by implementing more
general deformations of Kerr, and will 
explore the general validity of Petroff-Ansorg inequality.


\end{document}